\documentclass[a4paper,fleqn]{cas-sc}

\usepackage[authoryear]{natbib}

\def\tsc#1{\csdef{#1}{\textsc{\lowercase{#1}}\xspace}}
\tsc{WGM}
\tsc{QE}

\begin{document}
\let\WriteBookmarks\relax
\def\floatpagepagefraction{1}
\def\textpagefraction{.001}

\shorttitle{Didymos thermal inertia}    

\shortauthors{Novakovi\'c and Fenucci}  

\title [mode = title]{ASTERIA - Thermal Inertia Evaluation of asteroid Didymos}


\author[1,2]{Bojan Novakovi\'c}[orcid=0000-0001-6349-6881]

\cormark[1]

\affiliation[1]{organization={Instituto de Astrofísica de Andalucía (IAA-CSIC)},
            addressline={Glorieta de la Astronomía s/n.}, 
            city={Granada},
            postcode={18008}, 
            country={Spain}}
            
\affiliation[2]{organization={Department of Astronomy, Faculty of Mathematics, University of Belgrade},
            addressline={Studentski trg 16}, 
            city={Belgrade},
            postcode={11000}, 
            country={Serbia}}        

\author[3,4]{Marco Fenucci}[orcid=0000-0002-7058-0413]

\affiliation[3]{organization={ESA ESRIN / PDO / NEO Coordination Centre},
            addressline={Largo Galileo Galilei, 1}, 
            city={Frascati (RM)},
            postcode={00044}, 
            country={Italy}}
            
\affiliation[4]{organization={Elecnor Deimos},
            addressline={Via Giuseppe Verdi, 6}, 
            city={San Pietro Mosezzo (NO)},
            postcode={28060}, 
            country={Italy}}


\begin{abstract}
Asteroid Didymos, recently targeted by the NASA DART mission, is also planned
to be visited by the ESA Hera mission. The main goal of the DART mission was
to impact Dimorphos, the small satellite of Didymos, which was accomplished
in September 2022. This collision altered the Didymos-Dimorphos system,
generating a notable quantity of ejecta that turned Dimorphos into an active
asteroid, with some ejecta potentially settling on the surfaces of both
components. This prompts the investigation into the extent of post-impact
surface alterations on these bodies, compared to their original states.
The purpose of this study is to independently evaluate the pre-impact
thermal inertia of Didymos.
We employed ASTERIA, an alternative to conventional thermophysical
modeling, to estimate the surface thermal inertia of Didymos. The approach is based on a model-to-measurement comparison of the Yarkovsky effect-induced drift on the orbital semi-major axis.
These results, alongside existing literature, enable an evaluation of the impact-induced
alterations in Didymos's thermal inertia.
Our nominal estimate with a constant thermal inertia model stands at $\Gamma = 211_{-55}^{+81}$ J m$^{-2}$ K$^{-1}$ s$^{-1/2}$, while assuming it varies with the heliocentric distance with an exponent of $-0.75$
thermal inertia of Didymos is found to be $258_{-63}^{+94}$ J m$^{-2}$ K$^{-1}$ s$^{-1/2}$.  
Subsequent verification confirmed that this result is robust
against variations in unknown physical parameters. 
The thermal inertia estimates for Didymos align statistically with values
reported in the literature, derived from both pre- and post-impact data.
The forthcoming Hera mission will provide an opportunity to further corroborate these findings.
Additionally, our results support the hypothesis that the thermal inertia of near-Earth asteroids is generally lower than previously expected.

\end{abstract}


\maketitle

\section{Introduction}
\label{sec:intro}

The near-Earth binary asteroid (65803) Didymos was recently the target of
the NASA DART (Double Asteroid Redirection Test) mission
\citep{2018P&SS..157..104C}. The main aim of the mission, impacting Didymos's
accompanying body Dimorphos, was accomplished on 2022 September 26, 23:14 UTC.
The impact altered the binary system's orbital period from 11.92 hours to 11.37
hours, proving that kinetic impacts are a viable method of diverging an
asteroid orbit and, therefore, can also be applied for planetary defence
purposes \citep{2023Natur.616..443D,2023Natur.616..448T}.
For an overview of the achievements of the DART mission and related investigations we
direct readers to a recent paper
by \citep{Chabot_2024}.

Further enhancement of the ability to protect our planet crucially depends on
how well we understand the consequences of the DART impact experiment. For
instance, in the case of small objects such as Dimorphos, the impact
likely occurred in the strength-dominated regime, where the physical properties
of the surface play a significant role in the crater formation process
\citep[e.g.][]{1999Icar..142....5B}.

Thermal inertia gauges the resistance of a material to temperature changes and can
indicate the particle size of regolith
\citep[e.g.][]{gundlach-blum_2013,2024Icar..40715771S}, or the porosity of rocks
and boulders \citep{grott-etal_2019,2021NatAs...5..766S}. Therefore, it serves as a
valuable property that characterizes small bodies surfaces.

The impact on Dimorphos also generated ejecta, boosting the momentum by
a factor of 3.6 $\pm$ 0.2 \citep{2023Natur.616..457C} and effectively transforming
it into an artificial active asteroid
\citep{2023Natur.616..461G,2023Natur.616..452L}. A part of the ejecta may or
may not be deposited on Didymos \citep[see, e.g.][and references
therein]{2022PSJ.....3..118R,2022PSJ.....3..177F}. This raises the question of
whether a significant deposit of material at the surface of Didymos has
occurred, and if so, whether it changed the surface thermal inertia of 
Didymos. 

\subsection{Pre- and post-impact surface properties of Didymos}

\textit{What do we know about the pre- and post-impact Didymos' surface properties?}

Physical properties of the Didymos system before and after the DART impact were
studied by \citet{2023A&A...676A.116L}, using photometric observations. The
authors found variations in taxonomic classification from S-complex
(pre-impact) to C- and back to S-complex (post-impact). These variations are
likely caused by contamination from the ejecta. This implies that the properties
of the ejecta from Dimorphos could be somewhat different from those of the
material of Didymos surface. Therefore, accumulation of the ejecta on 
Didymos surface might change its properties.

\citet{2023arXiv231113483G} performed polarimetric observations of the
Didymos-Dimorphos system, spanning a period from about one month before to
almost four months after the impact. This allows the authors 
to determine the polarimetric behavior of the system in its original state
(pre-impact) and its changes following the impact of the DART spacecraft
(post-impact). \citeauthor{2023arXiv231113483G} found a drop in polarisation
in the post-impact measurements. They explained this as being, at least partly,
due to the ejection of smaller particles than those present at the surface
before the impact.

\citet{2023PSJ.....4..138M} analysed the ejecta dust properties and evolution
by applying Monte Carlo models to ground-based and Hubble Space Telescope observations 
of the Didymos-Dimorphos system, taken after the
impact. Among other results, \citeauthor{2023PSJ.....4..138M} show that up
to $1.5 \times 10^{6}$ kg of material fell back on the surfaces of both
Dimorphos and Didymos in the first 20 days following the impact. Some of this
material 
may have settled permanently on the surface of both binary system components.

Still, according to \citet{2023PSJ.....4..229P}, near-infrared spectral
observations of the Didymos system taken before the impact and after the ejecta
dissipated show no signs of the spectral features changes. The authors
suggested that both Didymos and Dimorphos are made of the same silicate
material, and interpreted this as a support for a binary asteroid formation
theory that includes the breaking up of a single body due to rotational
fission. \citeauthor{2023PSJ.....4..229P} also suggested that only a negligible
amount of non-weathered material was ejected from Dimorphos subsurface,
suggesting that Dimorphos originates from weathered material ejected from
Didymos surface.

The results described above significantly increase our understanding of the
Didymos-Dimorphos system. Still, these findings also leave some uncertainties
about which changes occurred on the surface of Didymos following the DART
impact on Dimorphos and, if they happened, to what degree. Better data and more
precise answers in this respect are expected from the forthcoming ESA Hera mission, which
aims to characterise the DART impact outcome \citep{2022PSJ.....3..160M}.

\subsection{Pre- and post-impact surface thermal inertia of Didymos}

The post-impact thermal inertia of Didymos has been recently estimated by
\citet{2023PSJ.....4..214R}, using data from the James Webb Space Telescope
obtained after the impact. Based on mid- and near-infrared observations, the
authors estimated the post-impact thermal inertia values at $260\pm30$ and
$290\pm50$ J m$^{-2}$ K$^{-1}$ s$^{-1/2}$, respectively.

The pre-impact thermal inertia of Didymos has also been recently estimated by
\citet{xxxx_PSJ2024}. The authors applied the standard procedure for
thermal inertia estimation, that is fitting thermal-infrared observations of a
planetary body or surface with an
appropriate thermophysical model \citep[e.g.][and references
therein]{delbo-etal_2015,2022PSJ.....3...56H}. Using the mid-infrared
observations obtained by ESO's Very Large Telescope before the DART impact,
\citeauthor{xxxx_PSJ2024} estimated the pre-impact thermal inertia of
Didymos to be $320\pm70$ J m$^{-2}$ K$^{-1}$ s$^{-1/2}$.

The statistical compatibility of the obtained pre- \citep{xxxx_PSJ2024} and
post-impact \citep{2023PSJ.....4..214R} thermal inertia values implies that
Dimorphos' ejecta likely had minimal impact on Didymos' thermal inertia.
Nonetheless, the slight shift towards higher pre-impact values warrants further
investigation into potential effects on thermal properties caused by ejecta.

In this letter, we contribute to the topic by taking advantage of an
alternative method for asteroid thermal inertia estimation recently proposed by
\citet{2024PSJ.....5...11N} to re-evaluate the pre-impact thermal inertia of Didymos.
By combining these findings with available literature values, we aim to place additional 
constraints on a possible difference between the pre- and post-impact thermal inertia. 

\section{Thermal inertia estimation by ASTERIA}
\label{sec:asteria}

The Asteroid Thermal Inertia Analyzer (ASTERIA) is a novel approach to asteroid
thermal inertia estimations. It is generally based on the model-to-measurement
comparison of the Yarkovsky effect-induced orbital drift. As such, ASTERIA is
independent from the classical thermophysical modeling approach, and it therefore
presents an alternative solution. A detailed description of the approach is
beyond the scope of this paper, and it can be found in \citet[][see also
\citet{2021A&A...647A..61F,Fenucci_et_al_AA2023}]{2024PSJ.....5...11N}, while the
corresponding code is freely available \citep{asteria_zenodo}.

Note that the ASTERIA framework, fundamentally grounded in the analysis of the Yarkovsky
effect, is tailored for the study of single bodies. However, it can still be effectively used 
for binary systems.
In fact, the Yarkovsky effect of a binary asteroid measured from astrometry
corresponds to that acting on the center of mass of the system. 
On the other hand, if the mass ratio of the system is small enough ($\sim$0.01 for
the Dimorphos-Didymos system, assuming the two components have the same density), the
Yarkovsky-induced drift predominantly reflects the influence on the primary
body \citep{vokrouhlicky-etal_2015}. Therefore, it is safe to compare the modeled 
Yarkovsky effect acting on the primary body to the measured drift of the center
of mass of the system.
This was further substantiated by results
from \citet{2024PSJ.....5...11N}, demonstrating that the ASTERIA model retains
its efficacy in binary systems, effectively accounting for the Yarkovsky effect
on the primary and, by extension, to the system as a whole.

The ASTERIA model has been demonstrated to provide reliable thermal inertia
estimations if certain conditions are met. In particular, along with the orbital parameters,
Yarkovsky drift and rotation period, at least the albedo and diameter of an
object need to be known, as outlined in \citep{2024PSJ.....5...11N}.
In the case of Didymos, these requirements are not only met, but supplemented
with additional relevant data, making Didymos a suitable candidate for applying
the ASTERIA model.

To accurately estimate the thermal inertia of Didymos using ASTERIA, it is essential to select
appropriate input parameters and an appropriate model of the Yarkovsky effect. To this
purpose, we identified a nominal set of input parameters and a model that
are most suitable to get accurate estimate of Didymos' pre-impact
thermal inertia. Additionally, to ensure the robustness of our nominal results, we have 
conducted analyses using alternative sets of input parameters and models.

\subsection{Nominal input parameters of the model}
\label{ss:nominal_parameters}

In our nominal case, we used JPL's solution \#204 (see Table~\ref{tab:didymos_orb}), where the Yarkovsky effect parameter $A_2$ was determined with a signal-to-noise ratio (SNR) of 16.
Note that this solution includes astrometry taken after the impact, although our goal is to assess Didymos's pre-impact thermal inertia. Considering that the arc of post-impact observations used for orbit determination is too short to influence the Yarkovsky effect's detection significantly, we assume that the determined $A_2$ value is representative of the pre-impact Didymos orbit. Accordingly, changes in the $A_2$ parameter with respect to previous JPL solutions (see solution \#181, Table~\ref{tab:didymos_orb}) are due to the higher accuracy of the orbit solution \#204, which includes radar and occultation measurements.

Physical parameters of Didymos used as the nominal settings are
summarised in Table~\ref{tab:didymos}. Additionally, the model
non-sphericity parameter derived from the ratio of the ellipsoid axes is
determined to be 1.03. 
Note, however, that we used the density distribution of S-type asteroids 
\citep[see][Table 2]{2024PSJ.....5...11N} for the density, without using any determination from the known literature. In contrast, we analyze how the density value may affect the results in Sec.~\ref{sss:alternative_parameters}.

We include constraints in the obliquity according to the value established by \citet{Naidu_PSJ2024}. Though in our original ASTERIA model \citep{2024PSJ.....5...11N} the population-based distribution of 
obliquity is used as input, we consider this constraint to increase the accuracy of the thermal inertia estimate. 
For the heat capacity and emissivity, which are unmeasured on asteroid Didymos,
we based our values on analogous measurements from meteorites. We adopted a
heat capacity value of $C = 600$ J kg$^{-1}$ K$^{-1}$, drawing on findings by
\citet{2020MAPS...55E...1O}. For emissivity, we used a value of $\epsilon$
= 0.9, in line with \citet{xxxx_PSJ2024}.

While the absolute magnitude $H$ is listed among the physical parameters, it was not
employed in the model since they it is redundant when both albedo and size are
known. However, we included it for reference purposes.

\subsubsection{Thermal inertia estimates with the nominal settings}
\label{sss:nominal_results}

To determine the thermal inertia of Didymos with the nominal set of input
parameters, we employed two different Yarkovsky effect models implemented within ASTERIA. The
first model assumes a constant thermal inertia along Didymos’ orbit. However,
Didymos' orbital eccentricity value of $e \sim 0.38$ may be large enough to
cause thermal inertia variations due to temperature changes along the trajectory. 
To accommodate this potential variability, we introduced a second model that adjusts 
thermal inertia along the orbit in relation to the heliocentric distance. 

An important point to emphasize is that the thermal inertia values for Didymos reported here are derived from orbit-average computations, corresponding to the object's average heliocentric distance, which is approximately 1.6 AU. This distinction is irrelevant when assuming constant thermal inertia, as the results are considered independent of heliocentric distance. However, this detail becomes crucial in the variable thermal inertia model (see below). 

\subsubsection*{Constant thermal inertia model}

The ASTERIA is first used with the constant thermal inertia model and nominal settings explained above. 
In this case, the thermal inertia of Didymos is estimated at $\Gamma = 211_{-55}^{+81}$ J m$^{-2}$ K$^{-1}$ s$^{-1/2}$. 
We consider this value our nominal estimate for the constant thermal inertia model. 

The thermal inertia of asteroids, mainly those covered by particulate regolith, may be sensitive to temperature variations \citep{rozitis-etal_2018}. As \citet{2021Natur.598...49C} found that the surfaces of S-type asteroids can efficiently produce particulate regolith due to thermal cracking and micrometeorite bombardment, the model incorporating thermal inertia variable with heliocentric distance may be more suitable for the asteroid Didymos. Nevertheless, it is important to note that our estimate with the constant thermal inertia model is statistically consistent with the pre- and post-impact results reported by \citet{xxxx_PSJ2024} and \citet{2023PSJ.....4..214R}. 

\subsubsection*{Variable thermal inertia model}

The variable thermal inertia model has one additional free input parameter, namely, the exponent $\alpha$ of a scaling formula $\Gamma = \Gamma_0 r^{\alpha}$, where $\Gamma_0$ represents the baseline thermal inertia at a distance of 1 astronomical unit (au), and $r$ is the heliocentric distance expressed in au. 

Therefore, in what follows, for all the results obtained with the variable thermal inertia model, the $\Gamma_0$ value that represents the baseline thermal inertia at a distance of 1 au will be reported. 

The theoretically predicted value of $\alpha$ is $-0.75$ \citep{delbo-etal_2015}, and we formally accepted the corresponding estimate as our nominal result for the variable thermal inertia case.
Still, the thermal inertia variations with heliocentric distance could differ for different asteroids \citep{rozitis-etal_2018}. Therefore, it is not a priori clear what value of exponent $\alpha$ should be used. 
To account for the unknown exponent $\alpha$, we used several values from the interval between 0 and -2, based on the finding by \citet{rozitis-etal_2018}.  

\begin{figure}[!ht]
    \centering
    \includegraphics[width=0.48\textwidth, angle=-90]{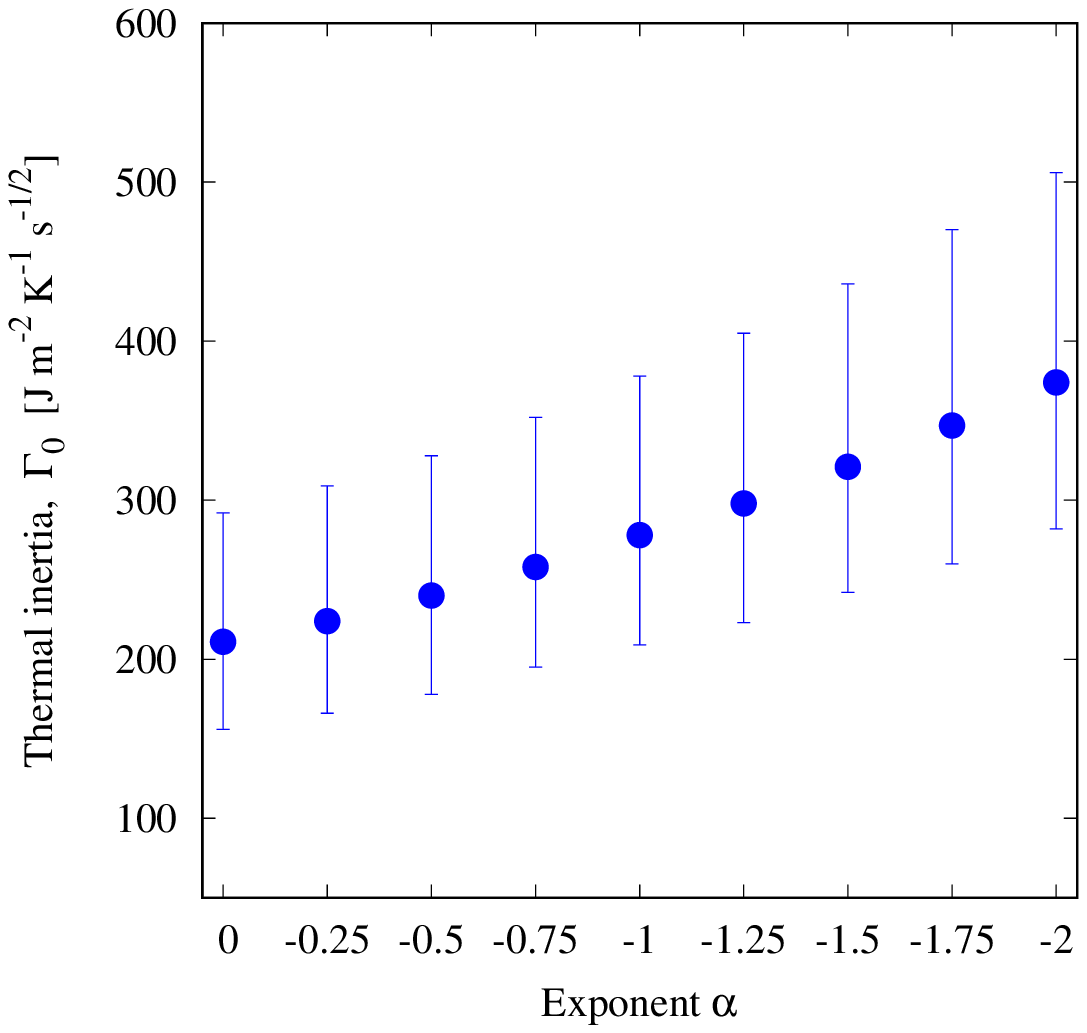}
    \caption{Thermal inertia estimates for asteroid Didymos with the model assuming thermal inertia variable with heliocentric distance. The results for different values of scaling exponent $\alpha$ are shown. Note that the $\alpha = 0$ case corresponds to the constant thermal inertia model, while the most commonly used value is $\alpha = -0.75$.}
    \label{fig:didymos_variable_TI}
\end{figure}

The obtained Didymos' thermal inertias are shown in Fig.~\ref{fig:didymos_variable_TI}. It should be noted here that the constant thermal inertia model corresponds to the $\alpha = 0$ case. The plot clearly reveals the sensitivity of the result to the exponent $\alpha$. The estimates range from $224_{-58}^{+85}$ J m$^{-2}$ K$^{-1}$ s$^{-1/2}$ (for $\alpha = -0.25$), to $374_{-92}^{+132}$ J m$^{-2}$ K$^{-1}$ s$^{-1/2}$ (for $\alpha = -2$). For $\alpha = -0.75$, the thermal inertia of Didymos is estimated to $258_{-63}^{+94}$ J m$^{-2}$ K$^{-1}$ s$^{-1/2}$.  

\subsubsection{Alternative input parameters}
\label{sss:alternative_parameters}

Although our results for nominal parameters align with the values reported in the literature, we also investigated how using other reasonable values of input parameters might potentially affect our conclusions. In particular, we tested different values for density, the magnitude of the Yarkovsky effect, heat capacity, and emissivity. Unless otherwise stated, we employed the constant thermal inertia model in all alternative parameter tests. In this case, the results should be compared to $\Gamma = 211_{-55}^{+81}$ J m$^{-2}$ K$^{-1}$ s$^{-1/2}$. However, when referring to a single result from the variable thermal inertia model, it denotes the baseline thermal inertia $\Gamma_0$ obtained with an exponent of $\alpha = -0.75$, and the reference value is $\Gamma_0 = 258_{-63}^{+94}$ J m$^{-2}$ K$^{-1}$ s$^{-1/2}$.

\subsubsection*{The role of density}

In our above analysis we considered the population based density of rocky S-type asteroids. 
While this is a reasonable assumption for Didymos, it is important to quantify how much this may 
affect the estimated thermal inertia.
Keeping all the parameters from the nominal model, but using the density value of
$\rho=2400\pm300$~kg m$^{-3}$ from \citet{2023Natur.616..443D} yields a
thermal inertia estimation of $207_{-61}^{+90}$ J m$^{-2}$ K$^{-1}$
s$^{-1/2}$. Similarly, using the density of $\rho=2790\pm280$~kg m$^{-3}$
from \citet{Chabot_2024, xxxx_PSJ2024}, results in a thermal inertia of
$203_{-57}^{+88}$ J m$^{-2}$ K$^{-1}$ s$^{-1/2}$. 

These values are statistically the same as those obtained with the nominal set of parameters, 
indicating that the estimated thermal inertia of Didymos is relatively insensitive to variations 
in the density, provided that their values are consistent with those of S-type asteroids.

\subsubsection*{The alternative values of heat capacity and emissivity}

We recall that the ASTERIA model is only weakly sensitive to the choice of the heat
capacity parameter, and using other reasonable values should not significantly change the
results. Nevertheless, \citeauthor{2020MAPS...55E...1O} also
found that the heat capacity increases with increasing temperatures, and
therefore it is higher for near-Earth asteroids. For this reason, we also
performed simulations assuming $C = 1000$ J kg$^{-1}$ K$^{-1}$.
Based on meteorite measurements, \citet{ostrowsky-bryson_2019} suggested
higher emissivity values than 0.9. Therefore, we also run simulations for 
emissivity of 0.984, as we used this value in our original ASTERIA model
\citep{2024PSJ.....5...11N}.

Using the larger heat capacity of $C = 1000$ J kg$^{-1}$ K$^{-1}$ did not
change our nominal result, while using the emissivity of $\epsilon$ = 0.984
marginally increased it to $215_{-56}^{+85}$ J m$^{-2}$ K$^{-1}$ s$^{-1/2}$.

\begin{table*}[]
    \caption{Physical parameters of asteroid (65803) Didymos.}
    \centering
    \begin{tabular}{lcc}
         \hline
         \hline
    Parameter & Value & Reference \\
    \hline
    Absolute magnitude, $H$  & 18.16 $\pm$ 0.06 & \citet{xxxx_PSJ2024} \\
    Diameter, $D$  & 730 $\pm$ 17 m & \citet{Barnouin_NatCo..2023} \\
    Best-fit ellipsoid, $a\times b\times c$  & $819\times801\times605$ m & \citet{Barnouin_NatCo..2023} \\
    Bulk density, $\rho$  & 2790 $\pm$ 140 kg m$^{-3}$  & \citet{Naidu_PSJ2024} \\
    Obliquity, $\gamma$  & 167.7 $\pm$ 0.5 deg & \citet{Naidu_PSJ2024} \\
    Rotation period, $P$  & 2.2600 $\pm$ 0.0001 hours & \citet{2023Natur.616..448T} \\
    Geometric albedo, $p_V$ & 0.17 $\pm$ 0.01 & \citet{xxxx_PSJ2024} \\
    Emissivity, $\epsilon$  & 0.9 & adopted \\
    Heat capacity, $C$   & 600 J kg$^{-1}$ K$^{-1}$ & \citet{2020MAPS...55E...1O} \\
         \hline
    \end{tabular}
    \label{tab:didymos}
\end{table*}

\begin{table*}[]
    \caption{Orbital parameters of asteroid (65803) Didymos.}
    \centering
    \begin{tabular}{lc}
         \hline
         \hline
          \multicolumn{1}{c}{} JPL solution 181 \\
          \hline
       Semi-major axis, $a$ & 1.644268882  $\pm$ $1.56 \times 10^{-9}$ au    \\
       Eccentricity, $e$ &    0.383882802  $\pm$ $2.75 \times 10^{-9}$    \\
       Inclination, $i$  &    3.407768167  $\pm$ $1.32 \times 10^{-6}$ deg    \\
       Parameter, $A_2$  & $(-1.8858 \pm 0.7226) \times 10^{-14}$ au d$^{-2}$  \\
    Observation Arc &  1996-Apr-11 to 2021-Feb-05 \\
    Epoch & 2457380.5 JD \\
     \hline
    \multicolumn{1}{c}{} JPL solution 204 \\
     \hline
       Semi-major axis, $a$ & 1.642665058 $\pm$ $2.72\times10^{-10}$ au   \\
       Eccentricity, $e$ &    0.383264789 $\pm$ $1.33\times10^{-10}$    \\
       Inclination, $i$  &    3.414150730 $\pm$ $1.61\times 10^{-8}$ deg    \\
       Parameter, $A_2$  & $(-1.0423 \pm 0.0649 )\times 10^{-14}$  au d$^{-2}$  \\
    Observation Arc & 1996-Apr-11 to 2023-Jan-21 \\
    Epoch  & 2460200.5 JD \\
       \hline
    \end{tabular}
    \label{tab:didymos_orb}
\end{table*}

\subsubsection*{The alternative orbit solution}
\label{sss:alternative_orbit}

We used JPl's orbit solution \#204 in our nominal case. However, as explained
in Section~\ref{ss:nominal_parameters}, strictly speaking, it also includes
post-impact observations, which means the resulting thermal inertia may not
purely reflect pre-impact conditions. We explored the potential outcomes using
an orbit solution based solely on pre-impact data to address this.

Accordingly, we employed JPL's orbit solution \#181
(Table~\ref{tab:didymos_orb}), which indicated a more substantial Yarkovsky
effect but had a significantly lower SNR. Our findings show
that solution \#181 leads to a higher estimate of Didymos' thermal inertia.
Specifically, when combined with the physical parameters from our nominal
settings, the thermal inertia values were calculated as $\Gamma =
304_{-99}^{+290}$ J m$^{-2}$ K$^{-1}$ s$^{-1/2}$ under a constant thermal
inertia model, and $357_{-98}^{+271}$ J m$^{-2}$ K$^{-1}$ s$^{-1/2}$ under a
variable thermal inertia model. 

It is crucial to highlight that the larger uncertainties associated with these
thermal inertia estimates directly result from the lower SNR 
in the Yarkovsky drift detected in JPL solution \#181. This emphasizes the
influence of data quality on the accuracy of our thermal inertia calculations.

\subsubsection*{Thermal inertia estimates with the minimal set of input parameters}
\label{sss:minimal_results}

In this part, we revisited the estimation of Didymos' thermal inertia employing
the minimal set of input parameters identified by
\citet{2024PSJ.....5...11N} as sufficient for ASTERIA to yield trustworthy
outcomes. The primary goal of this is to additionally validate ASTERIA's ability
to accurately estimate the thermal inertia with limited input data. Thus, the
significance of the results lies in their methodological implications rather,
than their specific values.
The minimum necessary set of input parameters for a reliable estimation is
obtained simply by replacing Didymos' obliquity value with the population-based
obliquity distribution \citep{tardioli-etal_2017}. In this analysis, we applied
both constant and variable models of ASTERIA's thermal inertia.

The constant model estimated the thermal inertia at $202_{-52}^{+79}$ J
m$^{-2}$ K$^{-1}$ s$^{-1/2}$, whereas the variable model yielded an estimate of
$255_{-57}^{+86}$ J m$^{-2}$ K$^{-1}$ s$^{-1/2}$. These results align with
those derived from the nominal set of parameters, reaffirming that the this
minimal set of input parameters suffices for ASTERIA to assess thermal inertia
reliably.

\section{Summary and Discussion}
\label{sec:summary}

Let us first briefly address the limitations of the ASTERIA model and the additional complexities related to the Yarkovsky effect that were not included in our analysis. Two primary factors in this regard are (i) the nonlinearity of boundary conditions for heat conduction at the surface and (ii) the anisotropy of thermal emission from the surface due to surface roughness ("thermal beaming").  

Nonlinearity effects tend to reduce the theoretically predicted semimajor axis drift rate da/dt within the relevant range of thermal inertia values, with a drop factor between 0.7 and 0.9 \citep{capek2007}. Conversely, thermal beaming effects generally increase the semimajor axis drift rate by a factor of 1.1 to 1.5 \citep[][see also \citet{2014PASJ...66...52M}]{2012MNRAS.423..367R}. Therefore, fortunately, these two effects tend to compensate for each other. Although some residual effects may persist in specific cases, their overall impact should be limited. In this respect, it is worth mentioning that, despite the modeling issues discussed above, the cross-validation of ASTERIA with thermophysical modeling results performed on ten well-characterized near-Earth asteroids plus asteroid Bennu shows full consistency of the results \citep{2024PSJ.....5...11N}. 

Another possible limitation is the dependence of the result of surface thermal inertia variations with latitude. Although this aspect can be included in the ASTERIA, the relevant information for Didymos is unavailable, preventing us from including it in the model. Nevertheless, suppose the level of potential variations is not much larger than in the case of asteroid Bennu \citep{2020SciA....6.3699R}, the impact on the result should be limited to a few percent \citep{2024PSJ.....5...11N}.

\begin{figure}[!ht]
    \centering
    \includegraphics[width=0.48\textwidth, angle=-90]{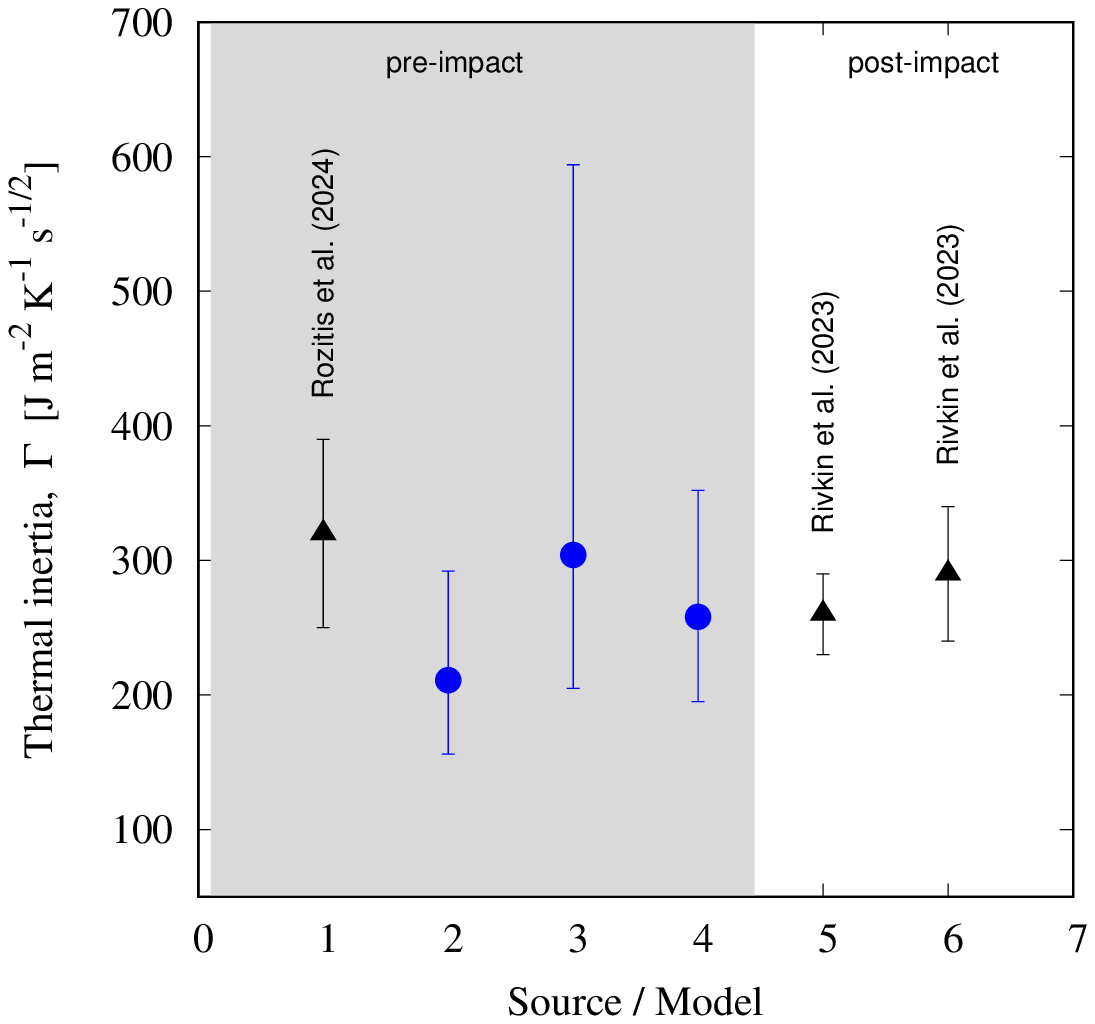}
    \caption{Thermal inertia estimates for asteroid Didymos. The blue circles
    show the results obtained in this work using the ASTERIA. The black
    triangles are literature values, as indicated in the plot. The plot is
    divided into two segments, corresponding to pre-impact
    and post-impact data assessments. The x-axis denotes the sources and models
    used for these estimations: 1) \citet{xxxx_PSJ2024},  2) ASTERIA (constant thermal inertia, orb. sol. \#204),  3) ASTERIA (constant thermal inertia, orb. sol. \#181),  4) ASTERIA (variable thermal inertia, $\alpha = -0.75$, orb. sol. \#204),  5) \citet{2023PSJ.....4..214R} (mid-infrared), and 6) \citet{2023PSJ.....4..214R} (near-infrared).    }
    \label{fig:didymos_TI}
\end{figure}

Our study aimed to obtain an independent estimate of the pre-impact thermal
inertia of the Didymos-Dimorphos binary asteroid system. The results of Didymos' 
thermal inertia estimates are summarized in Fig.~\ref{fig:didymos_TI}.
Our nominal analysis yields Didymos thermal inertia of $\Gamma = 211_{-55}^{+81}$, and $258_{-63}^{+94}$ J m$^{-2}$ K$^{-1}$ s$^{-1/2}$, obtained from the constant and the variable thermal inertia model, respectively. These estimates align well with the existing literature.

We also obtained results for different values of input parameters. 
We found that, as long as the input parameters are in a
reasonable range compatible with the known physical properties, they show
minimal impact on the resulting thermal inertia. 

The only exception is when an older orbit solution \#181 is used.
It yielded larger thermal inertia, more in line with results by \citet{xxxx_PSJ2024}, but with a significantly larger uncertainty due to a low SNR detection of the Yarkovsky effect. 

Specifically, increasing the non-gravitational acceleration parameter for a factor of about 1.8 resulted in thermal inertia of $\Gamma = 304$ J m$^{-2}$ K$^{-1}$ s$^{-1/2}$, more than 40\% larger than in the nominal orbit case. However, the latter estimate is associated with large uncertainties, mainly due to the considerable uncertainty in the non-gravitational acceleration parameter $A_2$, as determined from this orbit solution. 

The results obtained using the thermal inertia variable with heliocentric distance show a clear dependence on the exponent of the scaling relation. The resulting thermal inertia $\Gamma_{0}$ at 1 au is the lowest for exponent $\alpha = 0$ (equivalent to the constant thermal inertia model) and increases as the exponent goes from 0 to -2. This implies that, if the thermal inertia of Didymos changes with heliocentric distance, knowing the exponent $\alpha$ is crucial for proper estimation. 

On the other hand, supposing the thermal inertia estimates obtained by \citet{2023PSJ.....4..214R} are appropriate reference values, our results suggest that the thermal inertia of Didymos changes with an exponent $\alpha$ likely between $-0.75$ and $-1.25$. The upper bound aligns well with the value of $\alpha = -1.37 \pm 0.14$ found by \citet{2021PSJ.....2..161M}. This might further suggest that the larger estimate of $290\pm50$ J m$^{-2}$ K$^{-1}$ s$^{-1/2}$ by \citet{2023PSJ.....4..214R} is closer to the real thermal inertia of Didymos.

Technically, the thermal inertia estimates we obtained are statistically consistent with pre- and post-impact results as reported by \citet{xxxx_PSJ2024} and \citet{2023PSJ.....4..214R}. Combining all these results supports the hypothesis that the impact event on Dimorphos did not significantly affect Didymos' surface thermal inertia. Therefore any changes to Didymos' thermal inertia were either negligible or within the detection limits of current terrestrial observational techniques. However, further verification of these findings is required. This will likely be facilitated by the upcoming ESA Hera mission, scheduled to visit the system in late 2026.

In this respect, it is worth mentioning that thermophysical models tailored for binary systems such as Didymos-Dimorphos may further improve our knowledge of the system's thermal inertia. Such models are already under development \citep{2023arXiv230903458K}, but their full potential will be realized only once data from the Hera mission is available.  

In a broader context, our findings regarding the thermal inertia of Didymos
contribute to the accumulating evidence that many small near-Earth asteroids exhibit
lower-than-expected thermal inertia. For example, the global thermal inertia of
(162173) Ryugu was estimated to be $\Gamma = 225 \pm 45$ J m$^{-2}$ K$^{-1}$
s$^{-1/2}$ \citep{shimaki-etal_2020}, and for asteroid (101955) Bennu, it was
approximately $\Gamma = 300 \pm 30$ J m$^{-2}$ K$^{-1}$ s$^{-1/2}$
\citep{2020SciA....6.3699R}. Moreover, Bennu was found to host boulders with
even lower thermal inertia \citep{2024PSJ.....5...92R}. A study by
\citet{2021A&A...647A..61F} reported significantly low thermal inertia of
$\Gamma \lesssim 150$ J m$^{-2}$ K$^{-1}$ s$^{-1/2}$ for the rapidly rotating
asteroid (499998) 2011 PT, and similar findings were observed for the small,
fast-rotating near-Earth asteroid 2016 GE1, with estimates around $\Gamma
\simeq 100$ J m$^{-2}$ K$^{-1}$ s$^{-1/2}$ \citep{Fenucci_et_al_AA2023}.
Preliminary data from the Hayabusa2 extended mission targeting asteroid 1998
KY26 also indicate very low thermal inertia \citep{2021EPSC...15..390P}. These
studies collectively suggest that many small near-Earth asteroids, regardless of
their taxonomic classification, tend to have low surface thermal inertia.

With increased rate of Yarkovsky detections \citep[see e.g.][]{2024ExA....57....4L}, and the development of automated procedures for detecting the Yarkovsky effect, such as the one at the ESA NEO Coordination Centre \citep{2023arXiv231110175F}, the applicability of the ASTERIA model will be extended to many new objects, opening new possibilities for asteroid thermal inertia estimations. As shown by \citet{2024PSJ.....5...11N}, this will be especially important for small objects, where required input data for thermophysical modeling is generally unavailable.

In the case of potential deflection or exploration missions, a reliable Yarkovsky detection would be necessary from both pre- and post-deflection orbit to enhance the possibility of detecting thermal inertia alternations using solely ASTERIA. To achieve this, we underscore the significance of performing pre- and post-deflection occultation measurements\footnote{In a slightly different context, the importance of occultation measurements was also outlined by \citet{Makadia_2024}.}, which provide high-precision astrometric positions, allowing Yarkovsky detection from shorter observational arcs.

\section*{Acknowledgements}
The authors appreciate the support from the Planetary Society STEP Grant, made possible by
the generosity of The Planetary Society' members. This reserch was also partly supported by 
the Science Fund of the Republic of Serbia, GRANT No 7453, Demystifying enigmatic visitors of 
the near-Earth region (ENIGMA). BN acknowledges support from the project
PID2021-126365NB-C21 (MCI/AEI/FEDER, UE) and from the Severo Ochoa grant CEX2021-001131-S
funded by MCI/AEI/10.13039/501100011033.

\bibliographystyle{cas-model2-names}

\bibliography{didymos}


\end{document}